# World Models and the Law

Claudio Novelli[a]

[a] Digital Ethics Center, Yale University, New Haven, CT, USA

**Abstract**. Current AI may have a model of how we use human language, but this does not translate into a reliable representation of physical reality. World models promise to fill this gap by learning how an environment changes over time and predicting the consequences of an action, and they are already used to steer robots and generate explorable worlds. This article asks how AI regulation in three major legal contexts, the European Union, California, and China, copes with the distinctive challenges they raise. I propose a definition and a taxonomy of world models based on who or what acts on their predictions, and identify the risks of misplaced reliance on them. I argue that existing rules tend to cover world models but leave three questions open, i.e., action provenance, trajectory validity, and the weight of simulated evidence, and recommend ways to fill these gaps.

## 1. Introduction

Large Language Models (LLMs) are not built to have a model of the world: they may have a model of how we use human language, and perhaps fragments of knowledge about the world, but this does not translate into a reliable comprehension of the laws of physics or biology. For this reason, world models promise to be an alternative, or at least complementary, and groundbreaking technological paradigm, and they are gaining more and more credibility on the market.

From the legal perspective, there is little doubt that world models can fall under the scope of existing AI regulatory frameworks, depending on how they are used, but it is less obvious that these frameworks are ready for handling world models or if, by contrast, they require an adjustment effort comparable to the one triggered by LLMs (e.g., in the AI Act).[1] In this article, I will investigate whether some of the most prominent AI regulatory models are adequate for addressing the challenge posed by world models.

I try to show that what matters legally about world models is mostly about who, to what extent, what they generate is viewed as evidence of some kind. I stress the importance of the role played by world models in a workflow and the importance of handling the evidence generated in three ways: (a) by setting a taxonomy that sorts world models by who or what acts on their output (Section 3); (b) by identifying three risks that describe how such reliance can be misplaced (Section 4); and (c) by formulating seven recommendations for regulators (Section 6), all inspired by the principle that the weight given to a world model should be calibrated to the evidence that supports it. Some of these recommendations stem from an overarching legal concern: world models' outputs are claims about what an action would cause, and these claims unfold over a complex, interconnected sequence of events. Their reliability should therefore be checked over the span of time (Section 6.2), and in the real-world conditions (the envelope) (Section 6.3), in which a system relies on them. Others concern the data from which world models learn, and in particular how the actions in those data were chosen and identified (Section 6.1).

[1] Natali Helberger and Nicholas Diakopoulos, 'ChatGPT and the AI Act' (2023) 12(1) Internet Policy Review <https://doi.org/10.14763/2023.1.1682>, arguing that the risk-based approach of the Act had been written for systems with an intended purpose and fitted generative AI poorly.

## 2. What World Models Are

Cognitive scientists, researchers in reinforcement learning, the designers of video-generation systems, but also the marketing departments of companies selling three-dimensional content tools, have all used the term "world model". What they mean by it is not quite the same thing in each mouth. Let's try to narrow the scope to catch the technology we are interested in.

### 2.1. The roots

The very idea that entities in our reality, such as living organisms, carry a "small-scale model" of external reality in their heads and use it to try out alternatives before acting was first set out by the psychologist Kenneth Craik in 1943.[2] But after him, this idea, or similar variations, has been used across different branches, and some family resemblances can even be found in more normative or linguistic research areas (e.g., in cognitive linguistics, where "bachelor" presupposes an idealized model of marriage). The notion migrated into early AI mostly to give machines explicit, hand-coded models of their surroundings, which worked only in narrow, rule-governed micro-worlds.[3]

The more modern meaning of world models, the one that I specifically addressed here, dates from 2018, when David Ha and Jürgen Schmidhuber published a short paper simply titled "World Models". They use the case, pretty familiar in the heuristics literature, of a baseball batter, who has milliseconds to decide how to swing, less time than it takes visual signals to reach the brain. They stress that the swing is possible because the batter's brain predicts where the ball will be, and not by doing a precise calculus (but, again, relying on a heuristic). On this basis, they develop a system that compresses what an agent saw into a small internal representation, and learns to predict how that small representation would evolve in response to the agent's actions, and then trains the agent entirely inside its own environment - or what they call its "dream" of the environment - before transferring the result back to the real game.[4]

We can therefore try to give a first, approximative definition of a world model - in section 3 I will try to give a more operational one - as a system trained on observations of how an environment appears and transforms (e.g., video, images, sound, sensor readings) and, where available, on records of actions, with the consequences they produce. In this way, world models offer a tool that can be asked what will happen in a specific environment and, in most cases, what will happen if a given course of action is taken there. Eventually, they can be run forward to produce trajectories that were never observed because they are too rare, infeasible, expensive, or dangerous to explore.

In this working description, we can distinguish two features that we shall return to when addressing the governance, ethical, and legal relevant aspects that make world models particularly interesting:

**1) Most world models are action-conditioned.** "What usually comes next?" is a simple prediction, while "what would come next if this were done?" is a complex, conditioned one that requires a deeper representation of causality (and counterfactuality) that world model developers promise.[5]

---

[2] KJW Craik, *The Nature of Explanation* (Cambridge University Press 1943) ch 5.

[3] For the lineage from Craik through early AI to contemporary systems, see Stanford Institute for Human-Centered Artificial Intelligence (HAI), 'The World Model and Spatial Intelligence Era: Governing AI Beyond Language' (Issue Brief, 2026) 2–3.

[4] D Ha and J Schmidhuber, 'World Models' (arXiv preprint 1803.10122, 2018), published as 'Recurrent World Models Facilitate Policy Evolution' in *Advances in Neural Information Processing Systems 31* (2018).

**2) A world model can stand in for the environment it models.** Both software and hardware can be trained and tested within a world model (as in Ha and Schmidhuber's agent that learned inside its own dream), because the simulated environment serves as a surrogate for reality, where decisions can be rehearsed. This is a new form of simulation, based on learned regularities (i.e., inferred from visual data) and not on explicit rules. The rules a world model follows are implicit in the weights of a neural network, and, as I will argue in the next sections, this is one of the most distinctive governance and regulatory challenges posed by world models.

### 2.2. What world models bring that language models do not

The frameworks now in force were drafted with conventional AI (or large language models) in view, and the first objection to this article is that a world model is just a generative model in another modality. The stronger form of the objection is that language models already have world models, e.g., a system trained to predict text about the world may recover the world's structure, and probes have found a representation of the board inside a model trained only on Othello transcripts, and representations of geographical position and historical time inside large language models.[6] If so, the distinction is one of degree, and a regime that reaches language models reaches world models with it.

There are many reasons this objection is unconvincing and that invite thinking of world models as different in *kind* from language models. In the first place, and specifically regarding the stronger form of the objection, evidence from cognitive science shows that human world models are not linguistic in origin. We know that infants represent objects as solid and persisting in motion much earlier than they speak (persistence is, as Section 2.4 notes, the test now applied to artificial world models as well), and we carry this attitude for our whole life, to the point that a common experience as adults is to process physical scenes like an approximate internal simulation and not as a verbal inference. This happens because the brain's language network is functionally distinct from the systems that support world knowledge, as proved by the fact that people who lose language through aphasia can still reason about objects (or numbers). On this evidence, mastery of linguistic form, which language models plainly have, is separable from the knowledge of the world needed to use language, which is acquired through perception and action and must be supplied from outside the text.[7]

---

[5]Stanford HAI (n 3) 3: 'A central goal of world models is counterfactual reasoning: predicting the consequences of an action, including actions the model was never explicitly trained on.'

[6]K Li and others, 'Emergent World Representations' (International Conference on Learning Representations, 2023); W Gurnee and M Tegmark, 'Language Models Represent Space and Time' (International Conference on Learning Representations, 2024).

[7]ES Spelke and KD Kinzler, 'Core Knowledge' (2007) 10 Developmental Science 89; PW Battaglia, JB Hamrick and JB Tenenbaum, 'Simulation as an Engine of Physical Scene Understanding' (2013) 110 Proceedings of the National Academy of Sciences 18327; E Fedorenko, ST Piantadosi and EAF Gibson, 'Language Is Primarily a Tool for Communication Rather than Thought' (2024) 630 Nature 575; K Mahowald and others, 'Dissociating Language and Thought in Large Language Models' (2024) 28 Trends in Cognitive Sciences 517; S Harnad, 'The Symbol Grounding Problem' (1990) 42 Physica D 335; LW Barsalou, 'Grounded Cognition' (2008) 59 Annual Review of Psychology 617. Contra, M Abdou and others, 'Can Language Models Encode Perceptual Structure Without Grounding?' (Conference on Computational Natural Language Learning, 2021); E Pavlick, 'Symbols and Grounding in Large Language Models' (2023) 381 Philosophical Transactions of the Royal Society A 20220041. The argument shows not that language models carry no world structure, an empirical matter, but why what they carry would be partial; it rests on underdetermination, not anti-representationalism, and leaves the sense of 'world model' in Section 2.1 untouched.

This does not entail that a model of language has nothing to say about the world, and, to some extent, is a model of some parts of it, but we need to be careful about what such a model can depict. The philosophical ancestor of this result is, to some extent, the later Wittgenstein's rejection of the picture theory of his own Tractatus, on which the structure of propositions mirrors the structure of facts. The later Wittgenstein will claim that words gain their content from use within shared practices, not from mirroring the world.[8] There are good reasons to conclude that a model of language is therefore a model of how the world is described, and description underdetermines what it describes. We should bear well in mind that the same underdetermination afflicts models, whatever their use, trained on video alone. It is action data, not more footage, that separates a world model from a renderer.

On the "weaker" form of the objection, a relevant aspect that shows the difference between language and world models concerns then the evaluation standards: a language model predicts the next token, while a world model predicts the next state of an environment, typically given an action,[9] and is judged not on whether its output looks right but on whether the ball lands where it says it will.

Another key reason to view world models as a different beast is that a world model is expected to maintain a coherent state of the environment over time - and not just a sequence of tokens as a response to a given prompt - and what is out of view continues to exist. The representation of a responsive and persisting environment is learned from observation. It is not built in as in a video game, where every object is placed by a designer in advance.[10] Policymakers, as the Stanford HAI document highlights, must ask "[...] whether a learned environment matches reality closely enough to train a system on, test a system against, or guide a real-world decision".[11] Such difference is reflected in the governance of language models, where nothing really corresponds to this: we assess language models against reality, whereas a world model may be reality's stand-in, and it is from here that most complications arise.

Finally, and most obviously, the training data. Language models are trained on text, most of the time scraped from the internet and then fine-tuned, whereas world models are trained on video, imagery, three-dimensional scans, sensor streams and, crucially, on records of actions and their outcomes, which cannot be scraped at all but must be gathered by operating machines in real places.[12] This, as well, raises legal concerns, such as the data-protection profile (e.g., through continuous capture of real environments, faces, license plates, interiors and movement patterns). A different concern regards the most important data world models use, sometimes called action data. Since they are scarce, developers often compensate by pre-training their models on recordings that carry no action at all, and only later re-adapt them, e.g., V-JEPA 2 was pre-trained on over a million hours of internet video and then adapted with under 62 hours of robot interaction data (Section 2.4). Other times, developers let the model infer actions from the footage itself (as Genie did when it learned to respond to a player's inputs from game videos where no input was recorded). However, as I will argue in Section 6.1, this affects how far a world model can be trusted to predict what follows from an action.

---

[8]L Wittgenstein, *Tractatus Logico-Philosophicus* (1921) 2.1–2.19, 4.01; L Wittgenstein, *Philosophical Investigations* (1953) §§ 23, 43; for the form–meaning distinction in computational linguistics, EM Bender and A Koller, 'Climbing towards NLU: On Meaning, Form, and Understanding in the Age of Data' (Annual Meeting of the Association for Computational Linguistics, 2020).

[9]Stanford HAI (n 3) 3; Pierson Ferdinand LLP, 'Beyond LLMs: Why AI World Models Need Their Own Legal Playbook' (Client Alert, 29 October 2026).

[10]Stanford HAI (n 3) 3.

[11]ibid 2.

[12]Pierson Ferdinand (n 10) 'Differences in Training'; Stanford HAI (n 3) 12, noting that 'the scarce action-labeled interaction data cannot simply be scraped from the internet' and that this 'creates a steep barrier to entry'.

### 2.3. World Models Among Other Simulators

World models learn predictive representations of an environment from data and can use them to simulate how that environment might evolve. In this sense, they can serve as learned simulators or as components of larger simulations or decision-making systems.[13] World models therefore belong to computer simulation, an established field used for half a century to design aircraft, forecast weather, test drugs, plan cities, and reconstruct accidents, and whose epistemology, ethics, and evidential status have been examined at length.[14] Within this field, we should compare them with neighboring approaches - i.e., explicitly programmed simulators and physics engines, digital models, shadows and twins, and synthetic populations and agent-based models.

The first comparison concerns the *specification of dynamics*. In an explicitly programmed simulator, the modeler supplies algorithms and/or behavioral rules that determine how a certain simulated state, or a group of states, changes. As already mentioned several times at this point, what happens instead in a world model is that predictive mapping is learned from data. Developers still choose the training objectives and some architectural constraints, making modeling assumptions a core feature of world models.

The second comparison concerns the *relationship between a model and an identified target system*. Under one commonly used taxonomy, a digital model has no automated data exchange with its physical counterpart, while a digital shadow receives automated updates from it, and, at the highest level of connection, a digital twin has bidirectional integration.[15] A digital twin can incorporate learned predictive components, and a world model can be trained for a particular environment or used within a twin. By contrast, a world model need not maintain a live connection to any particular physical system, often because there's no physical system to connect to. Neither arrangement settles the question of predictive validity, as receiving operational data does not, by itself, validate a twin, nor does the absence of a continuously connected counterpart make a world model untestable. So, for example, researchers can, and it is frequently advisable, test a world model by comparing its predictions with observations that were not used to train it. Where feasible, it is recommended to test whether a strategy learned in simulation works in the environment the model represents: e.g., an agent trained to play a game inside a learned simulation can then be tested in the original game.[16]

The third and last comparison concerns how individuals and their interactions are *represented*. Known simulation techniques do several things, and world models represent something different and, at the same time, somehow gather them. So, take a synthetic population: it typically specifies who is represented, like artificial individuals or households whose attributes reproduce selected population statistics for a specific goal. But it does not, by itself, describe what they do, which is instead the aim of the simulation.[17] An agent-based model, instead, represents agents separately and simulates how they act and interact over time, assigning to them behaviors that may be programmed, learned, or partly both. One famous and interesting experiment is the one done by Park and others, in which generative agents combine a language model with explicitly designed memory and planning mechanisms, resulting in individual actions (and conversations) as well as emerging collective behavior.[18] A world model does not really represent each individual as a separate agent, because it learns how an environment changes, often *in response to actions*, to base its predictions.

---

[13] Ha and Schmidhuber (n 5); D Hafner and others, 'Dream to Control: Learning Behaviors by Latent Imagination' (International Conference on Learning Representations, 2020).

[14] E Winsberg, *Science in the Age of Computer Simulation* (University of Chicago Press 2010); WL Oberkampf and CJ Roy, *Verification and Validation in Scientific Computing* (Cambridge University Press 2010); on the ethics of modelling and simulation, FL Shults, WJ Wildman and V Dignum, 'The Ethics of Computer Modeling and Simulation' in *Proceedings of the 2018 Winter Simulation Conference* (2018) 4069; and, earlier, JP Sullins, 'Ethics and Artificial Life: From Modeling to Moral Agents' (2005) 7 Ethics and Information Technology 139, on simulation as 'a powerful medium for realizing complex thought experiments'.

[15] W Kritzinger and others, 'Digital Twin in Manufacturing: A Categorical Literature Review and Classification' (2018) 51(11) IFAC-PapersOnLine 1016, where the classification originates; A Fuller and others, 'Digital Twin: Enabling Technologies, Challenges and Open Research' (2020) 8 IEEE Access 108952.

[16] Ha and Schmidhuber (n 5); Hafner and others (n 14).

Video-generating world models illustrate how an environment can be represented without a separately specified agent model for each depicted entity. Genie, for example, generates action-controllable visual environments.[19] Video is only one possible output of a world model, however. Dreamer uses imagined trajectories in a compact latent state space to learn behavior.[20] In any case, when generated scenes depict people, their visual plausibility should not be mistaken for evidence that the model predicts how people would respond to a policy or other intervention. Such a claim requires validation using observations and measures suited to the behavior being predicted.

### 2.4. Applications

The most advanced application of world models is robotics, where it takes two main forms. In the first form, the robot *plans* within the model: before moving, it predicts the outcomes of several candidate movements and executes the one whose predicted result most closely matches the goal. The clearest published case is Meta's V-JEPA 2 (2025), which descends from the proposal by Yann LeCun (2022) that a model should predict in an abstract representation of a scene rather than pixel-by-pixel.[21] Something to bear in mind is that V-JEPA 2 was pre-trained on over a million hours of internet video and then, with not even 62 hours of robot interaction data, was given planned pick-and-place tasks on robot arms in two laboratories it had never seen. All without task-specific training or reward.[22] In the second form, the world model *manufactures training data*, which means that a robot learns actions, e.g., how to grasp or walk, from examples, but since collecting examples in the real world is slow and costly, the world model is used to generate thousands of plausible variations from a single recorded example (e.g., lighting and surroundings). This is roughly the business of NVIDIA's Cosmos platform, released as open models in January 2025 and expanded since, which NVIDIA says is used by 1X, Agility Robotics, Figure AI, and Skild AI to train humanoid and general-purpose robots.[23] Those descriptions come from NVIDIA; none of the companies has published results. What the world model produces in this use, and in the testing use described below, goes under the name, at least in legal scholarship, of synthetic data, i.e., "artificial data, generated by computer simulations or algorithms, which has analytical value".[24]

[17] C Dwork, K Greenewald and M Raghavan, 'Synthetic Census Data Generation via Multidimensional Multiset Sum' (arXiv preprint 2404.10095, 2024).

[18] JS Park and others, 'Generative Agents: Interactive Simulacra of Human Behavior' in *Proceedings of the 36th Annual ACM Symposium on User Interface Software and Technology* (2023).

[19] J Bruce and others, 'Genie: Generative Interactive Environments' in *Proceedings of the 41st International Conference on Machine Learning* (2024).

[20] Hafner and others (n 14).

[21] Y LeCun, 'A Path Towards Autonomous Machine Intelligence' (version 0.9.2, OpenReview, 27 June 2022).

[22] M Assran and others, 'V-JEPA 2: Self-Supervised Video Models Enable Understanding, Prediction and Planning' (arXiv preprint 2506.09985, 2025). This is still laboratory work: LeCun's company AMI Labs, which raised about one billion dollars in March 2026 to pursue the approach, had released no product as of November 2026 (company announcement of 10 March 2026, as reported in the press).

[23] NVIDIA, press releases of 6 January 2025 and 18 March 2025 on the Cosmos world foundation model platform.

Connected to robotics is autonomous driving, where the world model is more of a test bench: driving software is judged on how it handles rare events, and real driving produces too few of them to test against, so the model generates them. In February 2026, Waymo announced the Waymo World Model, built on Google DeepMind's Genie 3 and adapted to driving.[25] When this model is given a text prompt or a recorded drive, it produces camera images and lidar readings as a Waymo vehicle would perceive them in that scene, so the driving software can be run through it as if on the road. Waymo's stated purpose is to cover rare situations its fleet has never encountered, which is one of the most successful uses of world models, such as tornadoes, flooding, animals on the carriageway, and near-collisions, on the grounds that Genie 3's pre-training on a very large body of general video enables it to extrapolate to events absent from the company's own logs. On the public record, this is one of the first applications in which a learned environment is used to test a safety-critical system at commercial scale.

A third documented application is interactive content, where the world model produces an environment that someone can play in, or a three-dimensional world that someone can edit, e.g., for games, film and design. Of the applications considered here, this is the one closest to ordinary AI video generation, with OpenAI's Sora as the best-known example. Google DeepMind's Genie models learned, first from game footage and then from video across a wider range of subjects, to generate an environment that responds to the player's actions. With Genie 3 (August 2025), the player could move through the world as the model generated it in real time, and if the player wanted to change something in the scene, they could do so by typing while it ran. The scene held together for a few minutes, for the most part. Google DeepMind opened the model to subscribers in January 2026 under the name Project Genie, and in October 2026 released Genie 4, which generates the sound as well.[26] An example of the editing use is Marble, released in November 2025 by World Labs, the company founded by Fei-Fei Li. When Marble is given text, images, or video, what it produces is a three-dimensional scene that the user can then edit. [27]

There are further applications, but most are in earlier stages. For example, there are some attempts to couple language-model agents with world models or world-model-like planning modules, so that the agent rehearses the consequences of an action before taking it.[28] Elsewhere, like in engineering and the physical sciences, world models are proposed to test designs before they are built and to assess infrastructure under stress, an approach we have already seen in the use of digital twins or digital shadows. Furthermore, a related family of world models - they are world models only in the broad sense defined in Section 3 because they are more like structured scientific simulators, and not action-conditioned[29] - already forecasts the weather from data rather than from the equations of atmospheric physics.[30]

---

[24]Michal S Gal and Orla Lynskey, 'Synthetic Data: Legal Implications of the Data-Generation Revolution' (2024) 109 Iowa Law Review 1087; the definition is quoted here from I Cofone, KJ Strandburg and N Tilmes, 'Taxonomizing Synthetic Data for Law' (2025) 110 Iowa Law Review Online 217.

[25]Waymo, 'The Waymo World Model' (company blog, 6 February 2026).

[26]Google DeepMind, 'Genie 3: A New Frontier for World Models' (5 August 2025); Project Genie, announcement of 29 January 2026; Genie 4, research preview announced on 27 October 2026.

[27]World Labs, announcement of Marble (November 2025).

[28]M Parmar, 'Safety, Security, and Cognitive Risks in World Models' (arXiv preprint 2604.01346, 2026) s 3.3.

[29]Physical Intelligence Team, Shanghai AI Laboratory, 'A Definition and Roadmap for World Models' (arXiv preprint 2607.06401, 2026) s 2.6, describing a learned weather model as a structured scientific simulator that is not an agentic world model.

[30]R Lam and others, 'Learning Skillful Medium-Range Global Weather Forecasting' (2023) 382 Science 1416.

A further proposal concerns evidence, e.g., CCTV footage of a road accident. Kirfel and colleagues describe what they call a "counterfactual world simulation model", which would first use multimodal evidence of this kind to reconstruct what happened, and then to show what would have happened had the driver braked sooner or the pedestrian looked up. The authors have in mind use by investigators and also use in court, and set out an ethical framework for both.[31]

There are also applications in social simulation, probably the least developed of the uses described here, and in defense, where the examples given in the Stanford HAI brief are rehearsing a mission in a synthetic environment, managing logistics when communications are degraded, and analyzing cascading effects for intelligence purposes.[32]

Against all this, what should we conclude about the maturity of this technology? World models are already sold as products, some of them with open weights, and, as in the robotics and driving examples, they are already used as environments in which to train and test safety-critical systems. How reliable they are is harder to make out. The Stanford HAI brief, one of whose authors runs one of the leading companies, is quite candid that "a general-purpose world model capable of predicting physical consequences and supporting reliable planning and action across diverse real-world environments remains an active research frontier". In the early models, it did not take long for the model to lose track of a scene, e.g., objects would flicker or disappear, and one of the tests of how far things have improved is still whether an agent can move an object, leave the scene, and come back to find it where it was left.[33]

From a legal and governance perspective, the gap between what these systems are called and what they can do is itself worth analyzing, and there is little reason to wait for it to close. One reason is that technological maturity is a criterion used in flexible and time-dependent regulations. There is a further reason to keep an eye on this gap, which is that systems are already being described, sold and relied upon as models of the world in its complexity, while the evidence that they have learned the world's rules is still partial.

## 3. A Definition and a Taxonomy of World Models

Drawing together the features identified in Section 2, an operative definition of world models could be the following:

> 1. "World model" means a computational system, or an identifiable component of a system, that meets both of the following conditions:
>
> (a) it has learned from data, in whole or in substantial part, how environments change over time, including how the entities within them move, interact, and respond;
>
> (b) it is designed or adapted to use what it has learned to predict, from information about one or more actual or hypothetical states of an environment, one or more possible states that would follow them.

[31] L Kirfel and others, 'When AI Meets Counterfactuals: The Ethical Implications of Counterfactual World Simulation Models' (2025) 5 AI and Ethics 4593.

[32] Stanford HAI (n 3) 10–12.

[33] ibid 3.

Many world models are characterized by the fact that they show how the environment would change in response to a stimulus, which calls for a second definition:

> 2. "Action-conditioned world model" means a world model that can take an action as an input and predict how the environment would change if that action were taken, so that different actions can lead to different predictions.

In this definition, "environment" means a physical or virtual setting, including the entities within it, whose condition can change over time. "State" means any representation of the condition of the environment at a given time, including a sensor observation such as an image or video frame, or an internal representation that the system has learned and uses in its place. "Action" means an actual or hypothetical intervention in the environment by any agent, however represented, including a control signal, a learned representation of an action, or an instruction describing the intervention. Finally, a "prediction" may be a single outcome, several possible outcomes, or a probability distribution, and may be used solely within the system, including for comparison or selection of actions.

One of the goals of the definition is that it applies irrespective of the world model's architecture or learning technique, the form of its internal (latent) representations, or the medium of its inputs or outputs. It also applies whether it operates alone or within another system, the latter being the most common situation. Regarding its exclusionary criteria, a system falls outside the definition if the means it uses to predict environmental change are entirely determined by equations and parameter values specified by a person.[34]

### 3.1. A Taxonomy

Despite the technology's relative novelty, the literature offers several taxonomies, some of which already seem promising for circulation among regulators and policymakers.[35] Fei-Fei Li, for example, sorts world models by what they produce: renderers, simulators, and planners.

- A *renderer* generates observations, usually pictures or video, intended for human eyes, making visual plausibility the most important quality. To illustrate the distinctive character of a rendered output when compared to a "normal" AI-generated content, Li's own example is a video of a drone flying over a city: i.e., it may be flawless from above, but the buildings fall apart if you try to drive through them.[36]
- A *simulator* generates states, such as representations of a scene's geometry and dynamics, so that programs and professionals can compute on them. In this case, the most relevant quality is whether the scene "unfolds" correctly, and so, for instance, objects persist, and things fall as they should. To give an idea of the different use of a simulator when compared to a renderer, think of a crash-test model used by engineers rather than watched by an audience.

[34] Paragraph 1(a) is not satisfied where human-specified rules or equations wholly prescribe how the environment changes and the sole contribution of data, if any, is to estimate quantities designated in advance by those rules or equations as properties of the modelled environment. Human specification of the architecture or mathematical operations of a learning system is, by itself, insufficient to bring the system within this exclusion.

[35] X Li and others, 'A Comprehensive Survey on World Models for Embodied AI' (arXiv preprint 2510.16732, 2025), proposing a taxonomy by functionality (decision-coupled versus general-purpose), temporal modelling and spatial representation; J Ding and others, 'Understanding World or Predicting Future? A Comprehensive Survey of World Models' (arXiv preprint 2411.14499, 2024), dividing the field into models that build an internal representation to understand the world and models that predict its future states.

[36] F-F Li and the World Labs team, 'A Functional Taxonomy of World Models: Renderers, Simulators, Planners, and the Loop That Connects Them' (World Labs, 3 June 2026). The drone example is Li's own.

- A *planner* generates actions: given what it observes and a goal, it says what the agent should do next, and the quality that matters is whether the action achieves the goal.

This taxonomy gives an initial idea, but does not account for how the computation is performed within these systems. Just to give an idea, using it would be like putting an expert system and a neural network in the same category because they are both used for image recognition. This is one of the reasons that prompted alternative classifications, like the one from the Physical Intelligence Team at Shanghai AI Laboratory, which develops a two-dimensional taxonomy. In this alternative approach, Li's three functions are kept on one axis, and a second, architectural axis is added to show the way the model's knowledge of the world is represented within it.[37] The figure below illustrates how the two axes combine:

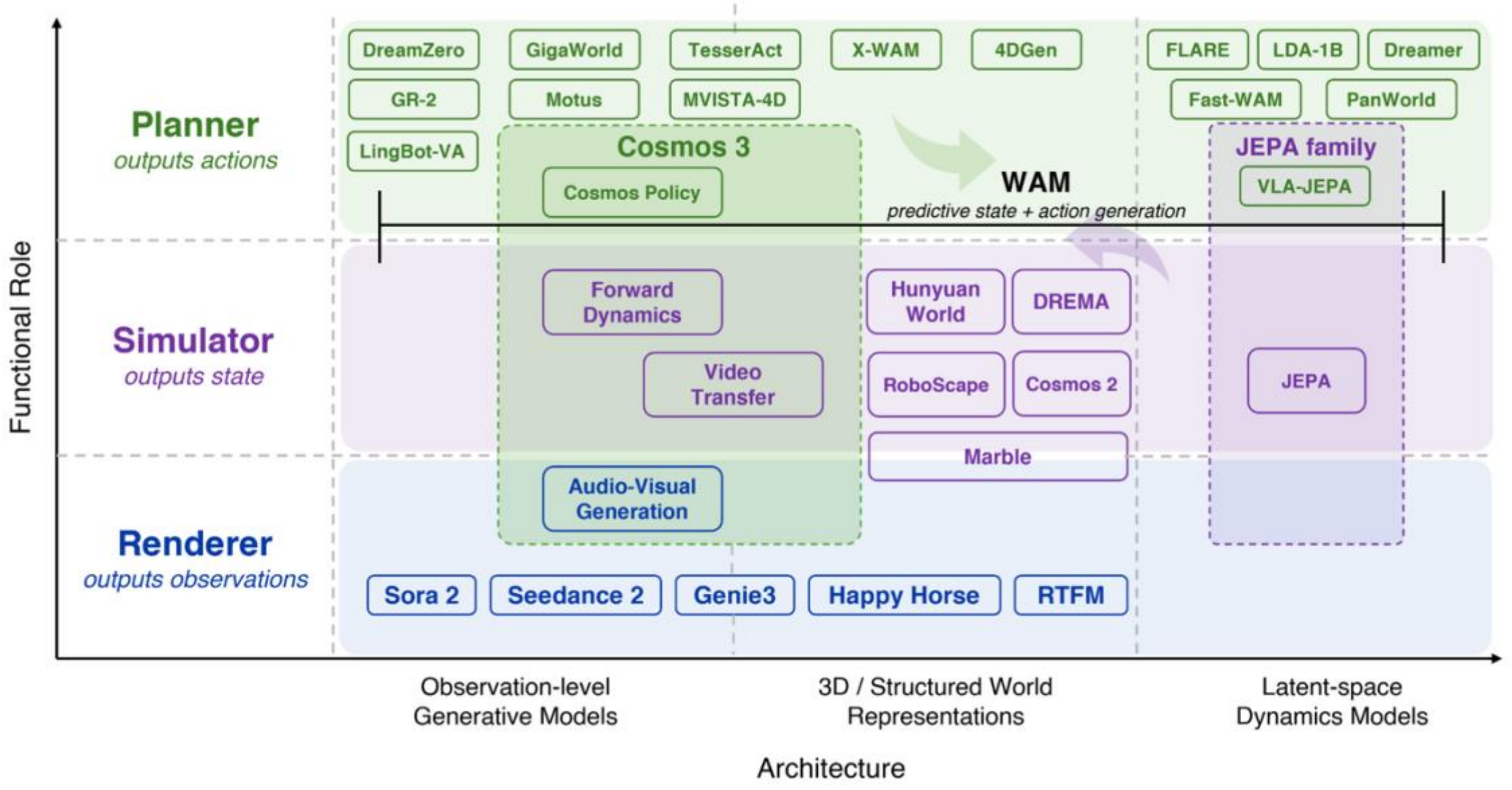

Figure 1. Dual-dimension taxonomy. Source: Physical Intelligence Team, Shanghai AI Laboratory, 'A Definition and Roadmap for World Models' (2026), fig 5.

The two axes represent a good starting point for our purpose because together they capture some technical profiles that matter for law: e.g., function determines the consumer of the output, and the consumer, in turn, determines or contributes to the legal regime. To give a basic idea: a renderer would probably be seen as a content technology, and this entails that its outputs raise the questions the law already asks of synthetic media, such as their provenance and the reproduction of real people and places. A simulator, instead, can be seen as a surrogate for (some portions of) reality, and consequently its outputs raise the many legally relevant questions, including the nature of the evidence generated: e.g., based on which criteria do we conclude that the surrogate is sufficiently faithful to what is being decided on its basis? Or even: what are the safeguards for avoiding transferring the simulated outputs directly into reality as if it were a real-world experiment? A planner, finally, can be treated as a control technology, bringing it into the busy space of legal issues raised by machines that act, i.e., product safety and liability for physical harm.[38] And the importance of the taxonomy does not just lie in capturing functions, because the world model's architecture also determines what can be disclosed and inspected (a pixel or a 3D scene can be viewed, whereas a latent state cannot). These are, in other words, aspects that a regulator, an auditor, or a certification body must look at.

[37] Shanghai AI Laboratory (n 32) ss 2.4–2.5: 'The first taxonomy therefore asks what a system does, whereas the second asks how the relevant world knowledge is represented and computed. They are complementary rather than competing, and a model should generally be assigned one or more labels along both axes.'

But I think that a different taxonomy should be considered for a governance perspective, one which is inspired by socio-technical literature, which yields a small number of types and then uses function and architecture to say what a regulator should expect inside each type and what to test for. In this sense, the sorting question is: who or what acts on the model's output? We may develop a taxonomy by distinguishing four roles that world models may perform in an application: control, advisory, surrogate, and content. Each type, i.e., each role that a world model plays in an application, comes with a typical function and architecture, but they are not mutually exclusive. Roles can overlap and, when this happens, the governance concerns should be considered together.[39] An illustration of my proposed taxonomy is the figure below:

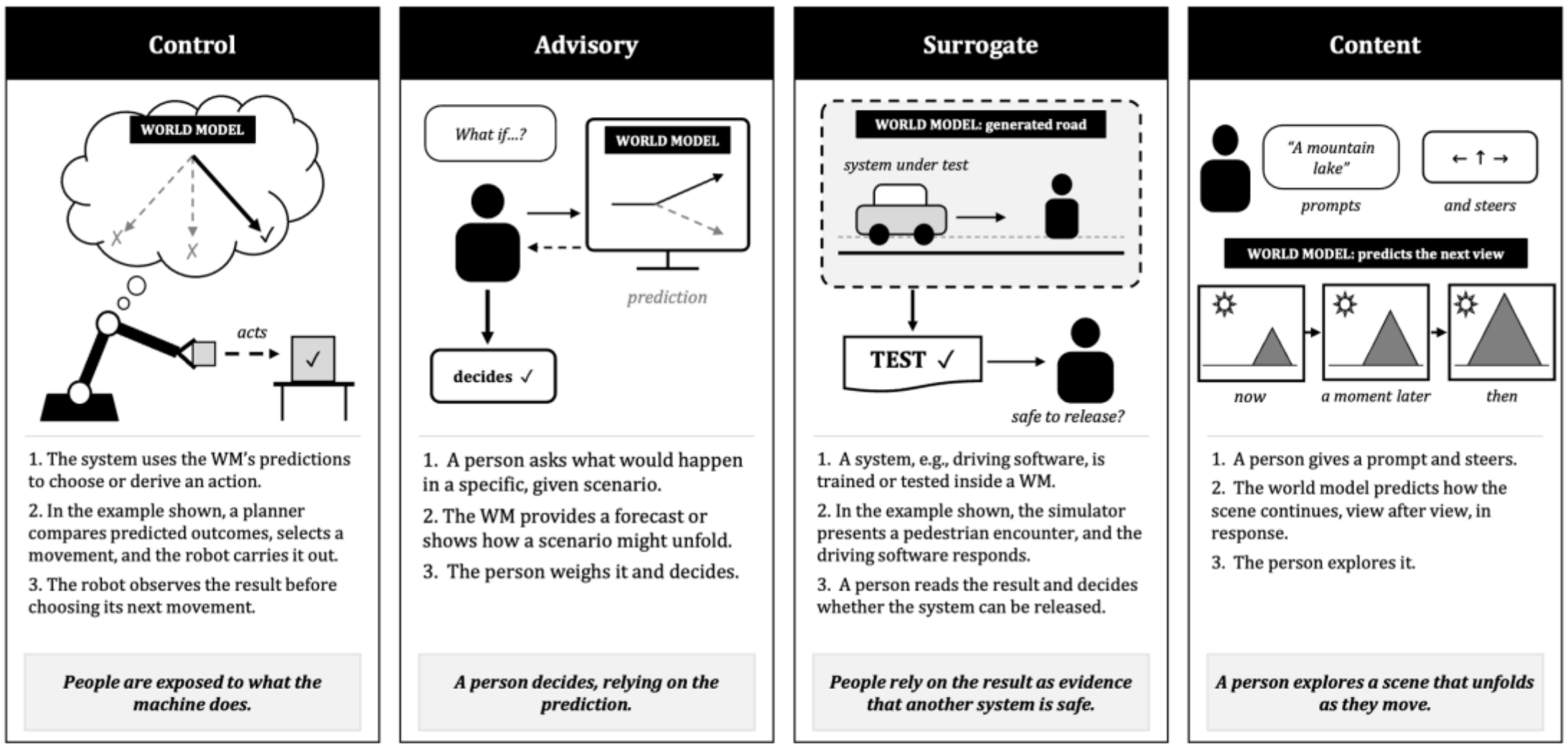

*Figure 2. Roles-based taxonomy*

1. **Control models**. The world model predicts the consequences of possible (courses of) actions in order to support automated action selection in a physical or virtual environment. An example is that of a robot's planner that uses predicted trajectories to pick movements, typically those with the best success rate, for placing a cup on a shelf. The architecture of control models can be latent-space, meaning the model predicts a compact internal state without having to produce an image, or observation-level, i.e., what the model predicts is what the robot would see if it made that movement.[40]
2. **Advisory models.** These models produce a prediction or a rehearsal for a person who decides, which can be a judge, an operator, a policymaker, or even the user of a software agent. Using the preceding classifications, their function would count as simulation, sometimes joined to planning. Their architecture is versatile, as it can be any of the three. Some examples, although mostly experimental, include the counterfactual reconstruction of a road accident for a court, an agent that rehearses an action before proposing it, and a simulator of how a population would respond to a policy. The learned weather forecasters mentioned in Section 2.4 also belong here, but, even though they inform human decisions, they do not predict the effect of an action.[41]

[38] Parmar (n 29) s 3.2 (scope note) on the different risk profiles of decision-coupled and general-purpose models, and s 8.4 on automation bias.

[39] A close taxonomic precedent, based on downstream use of world models, is in Bi and others, 'World Models: Predictive State, Generative Simulation, and Decision-Making Across Artificial Intelligence' (SSRN, 2026).

[40] Parmar (n 29) s 3.2 on the different risk profiles of decision-coupled and general-purpose models, Table 3 on the 'Foundry Problem' and s 8.4 on automation bias; Stanford HAI (n 3) 12 on the data barrier that favours those who already operate fleets.

3. **Surrogate models.** They stand in for the world to observe, train, and test another system within them. For this reason, their function is clearly simulation, and their architecture is observation-level or structured (3D). The generation of real-life (or rare) situations, such as the driving scenarios used by the Waymo World Model, falls into this category.
4. **Content models.** These models generate scenes for a person to view but, more importantly, to actively explore. The function is therefore a sophisticated, dynamic rendering, and its architecture is typically observation-level or structured (3D). Some content models for consumers include Project Genie and Marble, while a video generator such as Sora belongs here only insofar as it can predict how a scene develops.

Two caveats - or flags - can be added to this taxonomy. Some world models may have open weights or be licensed to many developers and therefore act as foundation (world) models, which should make us more cautious given risks such as propagation. A second question is whether the model is anchored to real places or people, which would make them more similar to the digital shadows and twins discussed in Section 2.3, and, of course, would raise privacy and surveillance risks (literature by Dirk Helbing and Novelli). A related problem, one that the literature on synthetic data highlights, once data are decoupled from the people and places they derive from, because they are fabricated rather than collected, consent and accountability are harder to trace.[42]

In any case, each type has a risk profile that would affect how laws address it, and the risk assessment would require careful consideration of its features (we cannot automatically transfer AI methodologies), but addressing this aspect goes beyond the scope of this article. The figure below summarizes some of the most significant aspects of these world model types.

## 4. The Risks of World Models

Before turning to the law, it is useful to state some of the most relevant risks posed by world models, and do so in the terms that AI regulation itself uses. The AI Act defines risk as "the combination of the probability of an occurrence of harm and the severity of that harm" (Art 3(2)) - which is also the conventional definition we find in many other documents and in many risk science articles - and asks providers of high-risk systems to identify the risks that are known or reasonably foreseeable (Art 9(2)(a)).[43] In this sense, the harms at stake with world models are familiar ones, i.e., injury to people, damage to property, and wrong decisions. What changes is the way these harms can arise, and how hard it is to foresee them. Let's try to name three relevant risks:

1. *Action-evidence risk.* I use this label, which, to my knowledge, is new, to denote the risk that a prediction about the effect of an action is relied on, even though the model has seen little or no reliable evidence about that action.[44]
2. *Exploitation risk.*[45] This is the risk that a machine or a person acts on a prediction that the model has overrated. Its source is the way predictions are used: each one builds on the previous one, so errors grow over a sequence,[46] and the action chosen is the one the model rates best, which tends to be the one it has rated too high.[47]
3. *False-assurance risk*[48]. This is the risk that a system is declared safe on the strength of tests or training carried out inside a world model that shares its gaps, or never learned the situation that matters. The harm here is indirect but wide-ranging, as it affects every system approved in this way, and even wider still when a single model is supplied to many developers.[49]

**5. A Comparative Analysis of World Models under Existing Laws**

[41]Kirfel and others (n 33), on the 'seeing is believing' effect.

[42]CD Whitney and J Norman, 'Real Risks of Fake Data: Synthetic Data, Diversity-Washing and Consent Circumvention' in *Proceedings of the 2024 ACM Conference on Fairness, Accountability, and Transparency* (2024).

[43]Claudio Novelli and others, 'AI Risk Assessment: A Scenario-Based, Proportional Methodology for the AI Act' (2024) 3 Digital Society 13. For a critique of risk regulation as an approach to AI, including that it works best on quantifiable problems and can present policy decisions as technical ones, see Margot E Kaminski, 'Regulating the Risks of AI' (2023) 103 Boston University Law Review 1347.

So far, the term "world model" appears explicitly in none of the legislation examined in this article, but only in standards projects and technical publications. One case is a Chinese national project titled "Artificial intelligence: World model evaluation standards", registered on 28 May 2026 on the national standards platform run by the State Administration for Market Regulation.[50] It belongs to the category of national standardization guidance technical documents, which offer technical guidance in fields that are still evolving and are not mandatory. Another case is IEEE P4216, a project to develop a Standard for the Evaluation of Artificial Intelligence World Models. The "P" in the name tells us that this is still a project, in the sense that the IEEE Standards Association has authorized a working group to draft the standard, but the standard does not exist yet. IEEE standards are also voluntary and are developed by consensus among industry, academic, and other experts. The project covers models that take in several kinds of input, such as images, sound, and sensor data, and that generate or predict how an environment changes. It aims to assess how consistent and realistic their outputs are, and how well those outputs support the tasks that other systems perform alongside them.[51]

Both projects aim to establish methods for evaluating world models, and such methods may help assess a model's performance, but compliance with a technical standard does not automatically establish compliance with the law. Its legal effect depends on whether, and under what conditions, the applicable legislation recognizes that standard.[52]

---

[44] Poor data representation or quality is, of course, not new to data science, but here the issue goes beyond whether the data reflect the setting of use and includes whether they show what follows from an action.

[45] Ha and Schmidhuber (n 5); T Kurutach and others, 'Model-Ensemble Trust-Region Policy Optimization' (International Conference on Learning Representations, 2018); M Janner and others, 'When to Trust Your Model: Model-Based Policy Optimization' in *Advances in Neural Information Processing Systems 32* (2019).

[46] E Talvitie, 'Self-Correcting Models for Model-Based Reinforcement Learning' in *Proceedings of the Thirty-First AAAI Conference on Artificial Intelligence* (2017).

[47] A Kamat and others, 'Learning the World: A Critical Survey of World Models for Physical AI' (preprint, 2026).

[48] F Liu and others, 'Security of World-Model-Based Embodied AI: A Lifecycle of Threats, Defenses, and Evaluation' (arXiv preprint 2607.28226, 2026). See also X Wang and others, 'Towards Safe World Models: Risks, Defenses, and Evaluation' (arXiv preprint 2610.13372, 2026).

[49] Parmar (n 29) Table 3.

[50] National Public Service Platform for Standards Information (State Administration for Market Regulation), project 20263035-Z-469, 人工智能 世界模型评测规范.

[51] IEEE Standards Association, P4216, Standard for the Evaluation of Artificial Intelligence World Models, project record. A further IEEE project that names world models, P4198, Guide for Application Architecture of World Models in Industrial Artificial Intelligence, appears in the draft minutes of the IEEE SA Standards Board of 25 September 2026.

The EU AI Act[53], for example, provides one instance of such recognition: under Article 40(1), a high-risk AI system or general-purpose AI model that conforms to a harmonized standard whose reference has been published in the EU's Official Journal is presumed to satisfy the legal requirements or obligations covered by that standard (which is not a presumption of compliance with the entire Act). Harmonized standards, however, are adopted by the European standardization organizations at the Commission's request[54]. Neither publication of a Chinese guidance document nor approval of an IEEE standard would, by itself, confer that status. Their content could gain this effect in the EU only indirectly, if a European standards body drew on them in a harmonized standard.

The rest of this section compares three jurisdictions, not because their rules are alike but because they attach duties to different objects, i.e., the European Union mainly to the use of a system and to the generality of a model, California to the developer of the largest models, and China to the service that is offered and, for vehicles, to the product.

## 5.1. The European Union

### 5.1.1. The AI Act

The AI Act (AIA) (Regulation (EU) 2024/1689, as amended by Regulation (EU) 2026/1744) regulates AI systems and general-purpose AI models. Its definition of an AI system, which is intentionally broad and inspired by a largely technology-neutral approach, covers all machine-based systems operating at varying levels of autonomy that infer outputs, including predictions, and are capable of influencing physical or virtual environments (Art 3(1)). Accordingly, a system built on a world model is likely to meet that definition. For the same reason, it would also fall under the OECD's definition of an AI system, which uses almost the same wording.[55] Whether the model itself is a regulated system, a general-purpose model, or a component requires a separate assessment instead. A second relevant question to understand if the AIA covers world models is whether it falls within the Act's territorial and material scope, which excludes certain research and development activities (Art 2(6) and (8)). A third is when the duties apply, given that the Act's provisions apply from different dates.[56]

The most important point, however, is that most duties in the Act depend on what a system is for. So, the intended purpose helps determine whether a system is classified as high-risk (Arts 3(12) and 6), whereas other provisions are triggered by capabilities, uses and outputs. This is why the four types of world models set out in Section 3 (which, as said, can overlap) correspond to four possible routes into the Act.

*Control models*. The route for this type of world model runs through product safety. An AI system can be high-risk when it is a product, or is intended as a safety component of a product, covered by Annex I legislation, and the product requires third-party conformity assessment (Art 6(1)). Such a system can include a world model that plans a robot's movements, even though controlling a machine is not enough, by itself, to satisfy either condition. From here on, everything depends on the product, because Annex I is divided into two lists which do not work in the same way. The first list (Section A) contains laws to which the Act adds its own requirements. If an AI system is a safety component of a product under one of these laws, the Act's rules for high-risk systems bind it directly, alongside the product's ordinary safety rules. The provider must then meet the requirements on data, human oversight, accuracy and robustness (Arts 10, 14–16), and the deployer must use the system according to its instructions and assign oversight to competent persons (Art 26(1)–(2)). The Commission must specify by delegated act, by 2 August 2027, which of these requirements may be limited where the sectoral law already gives equivalent or higher protection (Art 2(13)); no such act had been adopted at the time of writing. The law on medical devices is on this list (Annex I, Section A, point 11), so these duties would apply to a world model that guides a surgical robot. The second list (Section B), instead, contains laws that keep control of their own sector, such as those on vehicles and aircraft. For products under these laws the Act's requirements do not apply directly (Art 2(2)) and they take effect only once they have been written into the sectoral law itself (Arts 102–109). Machinery, the category that covers most industrial and service robots, was moved from the first list to the second in 2026 (Annex I, Section B, point 21), which means that a factory arm or a warehouse robot that plans its movements with a world model follows the sectoral route, the same one discussed below for vehicles.

---

[52] Ronit Justo-Hanani, 'Risk-Based Approach to EU AI Act: Benefits and Challenges of Co-Regulation' (2026) 9 Policy Design and Practice 171.

[53] Regulation (EU) 2024/1689 of the European Parliament and of the Council of 13 June 2024 laying down harmonised rules on artificial intelligence (Artificial Intelligence Act) [2024] OJ L, 2024/1689, as amended by Regulation (EU) 2026/1744 [2026] OJ L, 2026/1744 ('AI Act'). Unless otherwise indicated, the articles, recitals and annexes cited in Section 5.1 are those of the AI Act as amended.

[54] Regulation (EU) No 1025/2012 of the European Parliament and of the Council of 25 October 2012 on European standardisation [2012] OJ L316/12, art 2(1)(c).

[55] OECD, *Recommendation of the Council on Artificial Intelligence* (OECD/LEGAL/0449, adopted 22 May 2019, as amended).

[56] Under amended Article 113, Chapter III, Sections 1–3 generally apply from 2 December 2027 for Annex III systems and from 2 August 2028 for Annex I systems. General-purpose model obligations have applied since 2 August 2025, with a transition until 2 August 2027 for models placed on the market before that date (art 111(3)).

In short, the same system may be treated in two ways: i.e., a world model that controls a medical device is bound by the AI Act's requirements directly, while a world model that controls a machine or a vehicle, which is where the control models described in Section 2.4 are found, is reached only through the law of its sector, and only to the extent that this law has taken the Act's requirements on board. For a control model, the first question is therefore which list its product is on.

*Advisory models*. The route here runs through the Act's list of high-risk uses (Annex III). The most relevant entry of the AIA for these world models is point 8(a): systems intended to assist a judicial authority, or to act on its behalf, in researching and interpreting facts and law and applying the law to concrete facts, including in alternative dispute resolution. So, for instance, a system built to reconstruct an accident for a court may fall within it, on two conditions: that this is the purpose its provider intends, and that the exception in Article 6(3) does not apply. An image from a general-purpose tool does not acquire that purpose because someone files it in court. Other advisory uses may fall under other entries, such as employment, creditworthiness or access to public benefits. Once a system is high-risk, two safeguards address undue deference. Those who oversee it must be enabled to remain aware of automation bias (Art 14(4)(b)). And a person affected by a decision based on it may ask for an explanation of the system's role and the main elements of the decision, within limits: the right covers Annex III systems other than point 2, and decisions with legal or similarly significant effects that the person considers harmful to their health, safety or fundamental rights (Art 86(1)–(3)). Therefore, the Act governs how an advisory model is overseen and explained. Whether its simulation is admissible as evidence, and what weight it carries, will most likely be left to procedural and evidential law.[57]

[57] For an account, from the perspective of United States evidence law, of the validity and reliability to be established before evidence produced by AI is admitted, see Paul W Grimm, Maura R Grossman and Gordon V Cormack, 'Artificial Intelligence as Evidence' (2021) 19 Northwestern Journal of Technology and Intellectual Property 9.

*Surrogate models*. This type of world model has no clear route of its own. In fact, training or testing another system is not, by itself, a high-risk use under Article 6 and Annexes I and III. But the AIA may still reach surrogate models in three indirect ways. First, the same model may also do something the Act regulates: it may be installed in a product as a safety component, be intended for a listed high-risk use, or generate synthetic video, in which case the marking duty of Article 50(2) applies on its wording. A surrogate model may also qualify as a general-purpose AI model. That depends on how general the model is, not on its use for training or testing, so the conditions are set out separately below in the discussion of world models supplied to many developers. Second, its outputs may be caught by duties on whoever provides the high-risk system that is trained or tested with them.[58] These duties can require scrutiny of synthetic data and simulated tests. Third, the Act addresses the case in which the provider of a high-risk system relies on someone else's tool. Where a third party supplies an AI system, tool, service or component that is "used or integrated in" a high-risk system, the two must set out in a written agreement the information, technical access and other assistance that the provider needs in order to meet its own obligations under the Act (Art 25(4)). For a surrogate model, this would mean, for example, that a company which licenses a world model from another firm to generate its training data or test scenes could require that firm to disclose how the model was built and validated and what its known limits are. With some caveats: the rule concerns only tools obtained from a third party, so it does not touch a surrogate model that a company builds and uses in-house (the AIA may still apply, see above), and it does not apply to tools made publicly available under a free and open-source license, unless they are general-purpose AI models. In short, the Act has no rule for surrogate models as such, but it can reach them in the three ways just described: through duties on the model's own provider, where the model also does something the Act regulates; through duties on the provider of the system that is trained or tested with it; and through the agreement that must link the two where the model is supplied by a third party. Still, none of the three says how a learned environment is to be shown to be valid.
*Content models*. The route here runs through the Act's transparency rules, which impose two separate duties on two different actors. The first duty falls on the provider, the company that builds the system. If the system generates synthetic audio, images, videos, or text, the provider must ensure that the output carries a machine-readable mark and can be detected as artificially generated or manipulated (Art 50(2)). The second duty falls on the deployer, the person or organization that uses the system. If the content is a deepfake, the deployer must inform the audience that it was artificially generated or manipulated (Art 50(4)). In the AIA, a deepfake is an image, audio, or video that resembles existing persons, objects, places, entities, or events and would appear to be authentic (Art 3(60)). Such definition, therefore, includes places, which matters for world models: a generated scene that reproduces a real street or building closely enough to pass for footage of it would qualify.[59]

As anticipated, a widely distributed world model may also be a general-purpose AI model, which entails a separate set of duties that attach to the model itself rather than to any particular use. For this to happen, the model must display significant generality, perform a wide range of distinct tasks competently, and integrate into varied downstream systems (Art 3(63)). By contrast, the mere fact of being licensed to many developers, or released as open weights, does not show any of this by itself.

---

[58] That provider's training, validation and testing data must be relevant, sufficiently representative and, as far as possible, free of errors and complete for the intended setting (art 10(3)–(4)), and its testing must use metrics defined in advance and suited to the intended purpose (art 9(6)–(8)).

[59] There is an exception for law enforcement, and a lighter rule for works that are evidently artistic, creative or fictional, where it is enough to disclose the existence of generated content in a way that does not spoil the work.

But what does it mean for a system to be "general" according to the AIA? The response lies in the scale and generative capacity: models with at least a billion parameters, trained on large amounts of data via self-supervision, and large generative models that flexibly produce text, audio, images, or video (recitals 98–99). The Commission's non-binding guidelines of July 2025 turn this into a conventional working test: training compute greater than 10^23 FLOP, together with the ability to generate language (as text or audio), or images or video from text.[60] And the same guidelines give, as an example of a model outside the category, one trained specifically for modelling weather patterns or physical systems. For world models this produces a split. One that generates video from text prompts can qualify. One that predicts only in a latent space, or only from actions and sensor data, is unlikely to do so, however much computation it uses.

In any case, if a world model qualifies, two tiers of duties follow: the provider must keep technical documentation and give downstream providers the information they need to understand the model's capabilities and limitations (Art 53(1)(a)–(b)), and must have a copyright policy and publish a summary of its training content (Art 53(1)(c)–(d)). Models released under a free and open-source licence are exempt from the first pair of duties, unless they present systemic risk (Art 53(2)). The second tier covers only models with so-called "systemic risk", that is, models with high-impact capabilities. Such capabilities are presumed where training computation exceeds 10^25 floating-point operations, a presumption the provider may rebut, and the Commission may also designate a model (Arts 51–52). The provider must then evaluate the model, test it adversarially, assess and mitigate systemic risks, report serious incidents, and secure the model (Art 55(1)).

To conclude on this point, the AI Act has all the resources to reach a world model. Its definitions are broad enough to include one, and its duties attach to what a system is used for, what it outputs, and how general the model is. Control models enter through product safety, and for machinery and vehicles, where most of them are found, the Act's requirements arrive only through sectoral law. Advisory models enter through the list of high-risk uses, which brings rules on oversight and explanation but leaves the evidential value of a simulation to other law. Surrogate models have no door of their own and are reached through whatever else the model does and through the provider who trains or tests with it. Content models mainly enter through the transparency rules, which ask that output be identifiable. The general-purpose regime, finally, is keyed to language and text-prompted video, so it catches a world model by its generative face but may miss one that only predicts physical outcomes.

Across all of these, one of the most distinctive features of world models, a learned environment used in place of the real one, is met only by general requirements of representative data and adequate testing, and how such an environment is shown to be credible is a question the Act does not answer. The one EU instrument that addresses it was written for a single sector and is examined next.

### 5.1.2. The EU regulation on automated driving systems

Before a vehicle type can be sold in the EU, a public authority must certify that it meets the applicable technical requirements (this is called type-approval). The general safety requirements for vehicles are laid down in Regulation (EU) 2019/2144, which leaves the technical details to the European Commission, and Implementing Regulation (EU)

[60] Commission, 'Guidelines on the scope of the obligations for general-purpose AI models established by Regulation (EU) 2024/1689' C(2025) 5045 final (18 July 2025), para 17 and the examples that follow it.

2022/1426[61] is the Commission's rulebook for the case of vehicles that drive themselves without a human driver (fully automated vehicles).

Under the AI Act, the vehicle regulations (EU) 2018/858 and (EU) 2019/2144 are in the second list in Annex I of the Act (Section B, points 18 and 19), which lists sectors that retain control of their own rules. For AI systems in products on that list, only a few provisions of the Act apply (i.e., Art 2(2), as amended: Article 6(1), Article 60a and Articles 102–112; the provisions on regulatory sandboxes apply only once the high-risk requirements have been integrated into the sectoral legislation). This means that the Act's requirements on data, oversight, and robustness do not directly bind an automated driving system.

The object of the Implementing Regulation (EU) 2022/1426 is the automated driving system (ADS) of a vehicle type, and the duty lies with the manufacturer, which applies for approval, showing evidence that is then assessed by the type-approval authority, or by a technical service acting for it, against the specifications in Annex II and under the procedure in Annex III (2022/1426). Part 4 of Annex III, read here in the consolidated text of 24 March 2026, is the provision that matters most for us because it sets "sets "principles for credibility assessment for using virtual toolchain in ADS validation", and its stated aim is to demonstrate the credibility of "any simulation model or virtual toolchain" used to validate the driving system. Manufacturers test driving systems partly in simulation, a purpose a world model can serve, and the set of simulation tools they use is in fact what the Regulation calls a virtual toolchain. The manufacturer must also address which scenarios the simulation covers, where the models are known to be limited, how sensitive the results are to their assumptions, by what measures the models were validated, and how uncertain they are, including uncertainty about the model's structure (pts 3.3–3.4.5). In practice, the manufacturer must show that the simulated environment works, in part by documenting the data it used to verify that its simulation models match reality, and in part by demonstrating the calibration procedures by which it fitted the models' parameters to those data. The authority examines this documentation and may audit the manufacturer, either on paper or by conducting physical tests. The result of the assessment defines the tool's envelope, that is, the range of conditions within which its results may be relied on. How demanding the assessment is depends on how much the safety case rests on the tool (pt 3.3.5). And there is a clear floor to bear in mind, i.e., simulation cannot replace the physical tests that Part 3 requires (Part 2, pt 4.2).

This is the most explicit EU framework examined here for determining whether a simulated environment is sufficiently credible to validate a safety-critical system. Part 4 was written for simulation in general. It applies to "any simulation model or virtual toolchain" and makes no exception for one that learned its dynamics from data, in place of having them written by engineers. A manufacturer that validates its driving system with a learned world model would therefore fall under Part 4, but by interpretation: the provision does not mention such models, and it does not say how some of its requirements, such as explaining how a model's parameters were calibrated, are to be met by a model whose rules nobody wrote.

Against this background, EU vehicle law, unlike the AI Act, already contains a procedure for asking whether a simulation can be trusted, with documentation duties, an audit power, a defined range of validity and a floor of physical testing. It does so for one sector only, and it leaves the power to examine the evidence with the approving authority.

---

[61]Regulation (EU) 2019/2144 of the European Parliament and of the Council of 27 November 2019 on type-approval requirements for motor vehicles [2019] OJ L325/1; Commission Implementing Regulation (EU) 2022/1426 of 5 August 2022 laying down rules for the application of Regulation (EU) 2019/2144 as regards uniform procedures and technical specifications for the type-approval of the automated driving system (ADS) of fully automated vehicles [2022] OJ L221/1, cited from the consolidated version of 24 March 2026.

### 5.2. California

Notoriously, in the United States, the relevant rules have so far been made mainly by the states, despite many attempts by the federal government to do otherwise, and among them, California's statute is the most developed. It also resembles, mutatis mutandis, the EU AI Act, by which it was most likely influenced.

The Transparency in Frontier Artificial Intelligence Act (Senate Bill 53, 2025)[62], which took effect on 1 January 2026, does not regulate uses or products. It directly regulates the companies that build very large models and identifies them by the models they build, using mainly three labels. An "AI model" is a system that can infer from its input how to generate outputs "that can influence physical or virtual environments" (Business and Professions Code § 22757.11(b)), which is wording very close to the EU definition of an AI system. A "foundation model" is an AI model trained on a broad data set, designed for generality of output and adaptable to a wide range of distinctive tasks (§ 22757.11(f)). A "frontier model" is a foundation model trained with more than 10^26 integer or floating-point operations - ten times the threshold above which the AI Act presumes that a general-purpose model has high-impact capabilities - counting the original training and any later fine-tuning or other material modification by the developer (§ 22757.11(i)).

A world model would therefore be an "AI model" under the California Act and one trained on broad data and adaptable to many tasks would also be a "foundation model", whether or not it produces language. At the same time, fitting those definitions has little effect on its own, since the California Act imposes the principal governance duties (e.g., transparency and mandatory incident-reporting requirements) only on the developers of the subset of foundation models it calls "frontier models". Large frontier developers - i.e., those exceeding the statutory $500 million annual revenue threshold, including affiliates - must additionally implement, follow, and publish a catastrophic-risk framework, review it annually, and provide the additional assessment disclosures and internal-use reports (§§ 22757.11–22757.13).

The compute threshold, however, does not determine every obligation enacted by SB 53. If a world model qualifies as a foundation model, the accompanying whistleblower provisions can apply even when that particular model falls below the frontier threshold, provided its developer qualifies as a frontier developer through another model. The developer must protect qualifying employee disclosures concerning catastrophic risks involving the world model and notify covered employees of their rights; large frontier developers must also provide an anonymous internal reporting process (Labor Code §§ 1107–1107.1). These obligations do not themselves require publication of the non-frontier model's training compute or validation of its predictions.

For the principal model-governance duties, the question is whether a foundation world model's cumulative training computation exceeds 10^26 integer or floating-point operations. Public evidence is uneven, even if not totally absent. For instance, the original Genie paper reports 6.6 × 10^22 floating-point operations for training its dynamics model, below the threshold.[63] V-JEPA 2's reported training-resource requirements likewise support an estimate substantially below the threshold for the pretraining described.[64] These disclosures concern particular components or training stages and do not establish the complete cumulative computation relevant to statutory classification. Scaling experiments provide reasons to expect larger training runs in some world-model programs, but efficiency improvements and architectural differences prevent reliable predictions about whether, or when, particular models will exceed the threshold.

---

[62]Transparency in Frontier Artificial Intelligence Act, Cal Stats 2025, ch 138 (Senate Bill 53), adding Cal Bus & Prof Code §§ 22757.10–22757.16 and Cal Lab Code §§ 1107–1107.2. Section numbers cited in Section 5.2 without further indication are those of the Business and Professions Code.

[63]Bruce and others (n 20) Table 12.

[64]Assran and others (n 23) s 2.4 and fig 5.

Importantly, a distinct disclosure duty applies to covered genAI models or services that are made publicly available to Californians and mandates developers to declare whether synthetic data generation was used in development (California Civil Code § 3111(a)(12)). When and how synthetic data is deemed reliable is a different question the disclosure duty does not clarify. It may be relevant for that purpose to note that the nearest comparison with EU automated vehicle approval is California's staged testing and deployment permit process: a manufacturer seeking deployment must describe its completed safety case, along with the required evidence of testing (Cal. Code Regs., tit. 13, arts. 3.7–3.8, § 228.08). A world model used as an internal simulator can thus matter to a permit decision even when its developer falls outside SB 53.

### 5.3. China

Two particularly relevant routes through which Chinese law can reach world models are the regulation of generative content services and the regulation of automated vehicle safety.

China's framework for generative-content services rests on three agency-issued instruments. Each regulates a kind of service, not a kind of model, and each defines its service differently.

- The Provisions on the Administration of Deep Synthesis Internet Information Services (in effect since 10 January 2023) govern internet information services in China that use "deep-synthesis technology".[65]
- The Interim Measures for the Management of Generative AI Services (in effect since 15 August 2023) govern only generative services provided to the public in China (Art 2).[66, 67]
- The Measures for Labeling AI-Generated Synthetic Content (in effect since 1 September 2025) add labeling duties for providers already covered by the first two or by the rules on algorithmic recommendation.[68]

Because these rules attach to services, a world model is never covered simply for being a world model, but it is covered when two conditions are met: (i) someone uses it to provide a service of the kind an instrument describes, and (ii) what it produces is content of the kind the instrument names. This second condition is easily met by world models that generate scenes. Chinese law calls the generation or alteration of content with AI "deep synthesis", and Article 23 of the Provisions defines it to include virtual scenes, names techniques such as three-dimensional reconstruction and digital simulation, and describes immersive scenes that participants can experience or interact with. Therefore, an interactive, explorable environment generated by a world model fits this wording. Of course, the wording also catches systems that are not world models, since producing a virtual scene does not show that a system has learned how an environment changes. And it is unlikely to capture a world model that produces no content at all, such as one that only selects a robot's next movement.

[65] Provisions on the Administration of Deep Synthesis Internet Information Services (Cyberspace Administration of China, Ministry of Industry and Information Technology and Ministry of Public Security, Order No 12, in effect 10 January 2023) ('Deep Synthesis Provisions').

[66] Mimi Zou and Lu Zhang, 'Navigating China's Regulatory Approach to Generative Artificial Intelligence and Large Language Models' (2025) 1 Cambridge Forum on AI: Law and Governance e8, noting that the Measures do not cover research and development or enterprise and industry applications; see also Sara Migliorini, 'China's Interim Measures on Generative AI: Origin, Content and Significance' (2024) 53 Computer Law & Security Review 105985.

[67] Interim Measures for the Management of Generative Artificial Intelligence Services (Cyberspace Administration of China and others, Order No 15, in effect 15 August 2023) ('Interim Measures').

[68] Measures for Labeling AI-Generated Synthetic Content (Cyberspace Administration of China, Ministry of Industry and Information Technology, Ministry of Public Security and National Radio and Television Administration, issued 7 March 2025, in effect 1 September 2025) ('Labeling Measures').

The first condition (i), the service, decides which uses are covered. A company that offers the public an online tool for generating explorable worlds provides a service within all three instruments. By contrast, a company that builds a simulator and uses it only internally is outside the Interim Measures, which expressly leave out organizations that develop and apply generative artificial intelligence technology and do not provide generative artificial intelligence services to the domestic public (Art 2); whether it is also outside the other two instruments depends on their own terms.

Although the rules here apply to the service, their duties extend to the technology behind it in three ways.

- *Identification.* The scenes and other content that the model generates must bear a technical mark and, where the public could be confused or misled, a conspicuous label (Provisions, Arts 16–17). The Labeling Measures, in particular, specify two labels: an explicit one, such as a visible or audible notice, in prescribed circumstances, and an implicit one in the file's metadata. They also place duties on distribution platforms, application stores and users who publish. A provider may supply content without the explicit label at a user's request, if the law otherwise allows it, an agreement sets out the user's responsibilities and the logs are kept for at least six months; the metadata label is still required (Art 9).
- *Governance of the technology.* The provider must regularly review and verify its generative algorithms, which, where a world model powers the service, include that model, and certain tools require a safety assessment (Provisions, Art 15). Some duties also fall on the companies that supply the underlying technology.
- *Accuracy.* The provider must take effective measures to improve the reliability of generated content (Interim Measures, Art 4(5)) and the quality of training data (Art 7(4)), e.g., assessing the quality of data labeling and checking samples for accuracy (Art 8), and must give supervisors information about its training data and algorithms on request (Art 19).

These duties concern how content is identified and how a service is run. They include general duties to improve accuracy, but no procedure for showing that a learned simulation is valid for a particular safety-critical use. And they reach internal uses unevenly, as shown above. A world model used inside a company to train or test a safety-critical system is therefore the case this route handles least well.

The second route addresses that case for one sector. GB 44721-2026 is a mandatory national standard on the safety of automated driving systems, issued in the summer of 2026 and applicable from 1 July 2027. It regulates a product, the vehicle and its driving system, and it is binding: products and services that do not comply may not be produced, sold, imported, or provided (Standardization Law, Art 25).

The standard applies to vehicle manufacturers, who must demonstrate that their driving systems are safe. According to MIIT, they must do so through safety-assurance processes, safety documentation and simulation, test-track and road testing, with confirmatory tests by third-party bodies.[69] A world model comes within the standard in two ways: when a manufacturer builds it into the driving system itself, and when a manufacturer uses it as part of that simulation testing, that is, as a surrogate for the road.

[69]Ministry of Industry and Information Technology, statement on the release of GB 44721-2026 (4 August 2026), as reported by Xinhua (5 August 2026).

Therefore, seen through the four types of the taxonomy we have proposed in Section 3.1, Chinese law treats world models in different ways. A *content* model is regulated directly, as a service that generates content for the public: its output must be marked and labeled, and its provider must review its algorithms and take measures to improve accuracy.[70]

A *control* model is regulated as part of the product it steers. For vehicles, as we have seen, a mandatory standard requires the automated driving system to be shown safe.[71] For industrial robots, mandatory safety standards have long been in place, following international texts from 2006 and 2011 that concern the robot and its safeguards.[72] For humanoid robots, standards are still being planned, and the draft plan is not a mandatory national standard but lists world models as subjects to be standardized.[73] Behind all three stands the Product Quality Law, under which a product is defective if it fails to meet one of the applicable national or industry standards on personal or property safety or poses an unreasonable danger.[74]

An *advisory* model is regulated at the point of decision. In the courts, the results from AI shall be for supplemental reference only,[75] and the evidence rules test whether electronic material is genuine, with expert appraisal available where needed, without asking whether a simulation is correct.[76]

Finally, a *surrogate* model is regulated only where a product rule calls for simulation, which, from July 2027, will mean vehicles.[77] A simulator developed and used internally falls outside the Interim Measures; whether it also falls outside the other service rules depends on the terms of those rules.[78] In none of the instruments examined is there a general rule for demonstrating the validity of a learned environment.

## 6. What remains uncovered, and what adaptation would require

The preceding analysis clarified that world models would fall within the scope of major existing AI regulations, partly because world models are, like other machine-learning systems, systems that produce predictions, and partly because the rules are written in general terms and do not depend on a particular technology. A different question, though, concerns the adequacy of these frameworks.

I will identify where existing rules fall short for world models, but I will not advocate for dedicated regulation.[79] Let me start with the data on which world models are trained (Section 6.1), since the two problems that follow depend on them: the accuracy of world models as predictors (Section 6.2) and their validation as environments in which other systems are trained or tested (Section 6.3). At the end of each section, I will try to summarize specific recommendations, phrased in a sufficiently general way so as not to be limited to a particular jurisdiction.

---

[70] Deep Synthesis Provisions (n 67) arts 15–17 and 23; Labeling Measures (n 70) arts 3–5; Interim Measures (n 69) arts 4(5) and 7(4).

[71] GB 44721-2026 (n 71).

[72] GB 11291.1-2011, Robots for industrial environments: Safety requirements, Part 1: Robots (identical to ISO 10218-1:2006); GB 11291.2-2013, Part 2: Robot systems and integration (identical to ISO 10218-2:2011).

[73] Ministry of Industry and Information Technology, Guide for the Construction of the National Humanoid Robot Industry Standards System (2026 edition), draft for comment (24 August 2026).

[74] Product Quality Law of the People's Republic of China, art 46.

[75] Supreme People's Court, Opinions on Regulating and Strengthening the Applications of Artificial Intelligence in the Judicial Fields (8 December 2022).

[76] Supreme People's Court, Several Provisions on Evidence in Civil Procedures (as amended in 2019, in effect 1 May 2020), arts 14, 93 and 94.

[77] GB 44721-2026 (n 71).

[78] Interim Measures (n 69) art 2.

[79] Lyria Bennett Moses, 'How to Think about Law, Regulation and Technology: Problems with "Technology" as a Regulatory Target' (2013) 5 Law, Innovation and Technology 1.

One shared concern in the following analysis is that while most AI systems make one-off predictions that can be checked against observed cases, world models predict what an action would cause and how its effects unfold step by step. This is why some of the remedies proposed below are temporal in nature.

### 6.1. Action Data

It should be clear at this point that a world model typically lacks an explicit causal model that its developer could describe and disclose, since no one wrote its rules. One thing the developer can document is the evidence of how the model learned to associate actions with outcomes. But here a distinction matters for the legal reasoning: how the actions were chosen and how actions were identified. The distinction is familiar in causal inference and machine learning, but treating them as the content of a legal duty to document training data is, to my knowledge, new.

The first question distinguishes *experimental* from *observational* data. In an experiment, a researcher might intentionally and specifically instruct a vehicle to brake at different points on the same test track. If the vehicle starts at the same speed and the road conditions remain comparable, the researcher can investigate how delaying the command affects the stopping position. In this case, though, the strength of the evidence comes from the design of the comparison, and not from the mere fact that someone chose the action: instructing a vehicle to brake does not, by itself, show what braking caused. In observational data, such as routine fleet recordings, drivers choose when to brake for their own reasons, e.g., a driver may see a child near the road before the vehicle's camera records the child. The recording shows what happened after she slowed down, but estimating what would have happened had she kept her speed requires some (additional) account of the circumstances that made her slow down. In other words, observational data can support this kind of inference only when the relevant circumstances are adequately recorded, and comparable cases are available.[80]

The second question concerns how the action itself was identified. A vehicle log may record when the software issued a braking command, when the brakes engaged, or only how the vehicle moved, and these are three different pieces of information. If the log contains only video, a developer might *infer* braking from a drop in speed, although the car could also have slowed because the road ran uphill. Genie illustrates the problem in a virtual environment because it learned from game videos without access to the players' commands, and identified changes in the scene that could then be used to control it. This makes the generated scene interactive, but whether an inferred control corresponds to an action a player actually took is a further question that should not be presumed. Experiments on actions inferred in this way show that their quality drops when the footage contains changes correlated with the real actions but not caused by them, and that it improves markedly once recorded actions are supplied for even a small share of the data.[81]

This indicates what is worth documenting. For actions on which a safety claim relies, the developer should document how the actions were chosen, how they were recorded or inferred, and under what conditions they were observed.[82] Where the answer is unknown, as it often will be for video collected from the internet, the documentation should say so. An authority could then see, e.g., that a model was trained on many precisely recorded braking commands on dry roads but on very few on ice, and decide where additional testing is warranted. Whoever relies on a simulator to test a driving system should also disclose any overlap, known or reasonably ascertainable, between the data or pretrained models used by the driving system and those used by the simulator, where the overlap bears on the behavior being tested. If both learned braking behavior from the same fleet, a successful simulation may simply reproduce an error the two systems share; even if different fleets are used, the question may remain unanswered if neither regularly operates on ice.

[80]A related problem is documented for systems that learn to act from recorded demonstrations, which can mistake what merely accompanies an expert's action for its cause, including in driving settings: Pim de Haan, Dinesh Jayaraman and Sergey Levine, 'Causal Confusion in Imitation Learning' in *Advances in Neural Information Processing Systems 32* (2019).

[81]A Nikulin and others, 'Latent Action Learning Requires Supervision in the Presence of Distractors' in *Proceedings of the 42nd International Conference on Machine Learning* (PMLR 267, 2025) 46427.

Is this context about data legally covered? Partly. Under the AI Act, for instance, several provisions already provide a place for this information. Where the high-risk requirements apply, Article 10(2)(b) concerns how data were collected and where they came from, and Annex IV, point 2(d), calls for technical documentation on the provenance, selection and, where relevant, labeling of training data. Moreover, Annex IV, point 2(a), requires the provider to describe its recourse to pretrained systems or tools supplied by third parties and how they were used, which covers at least part of the overlap just mentioned. Comparable documentation duties apply to providers of general-purpose models under Article 53(1)(a) and Annex XI, Section 1, point 2(c), subject to the qualifications in Article 53. The gap hence lies in what these provisions leave unsaid, and guidance should add, that is, what a useful account of provenance looks like for an action-conditioned model: e.g., “fleet video” identifies a source but leaves the reader unable to tell whether the developer had access to braking commands or *inferred* them from the footage. In this sense, the proposal provides more precise content to an existing documentation duty where applicable, while extending it to every supplier of an internal simulator would require a separate legal basis.

In China and California, the gap is wider, as their data-disclosure rules do not reach this case. China’s Interim Measures require lawful sources of training data and, where data are annotated, rules for annotation and checks on its quality (Arts 7–8), but they concern generative services offered to the public in mainland China. California requires developers of generative AI systems or services made publicly available to Californians to publish a high-level description of their training datasets, including their sources or owners (Civil Code § 3111). A manufacturer that uses a simulator only to test its own vehicles ordinarily falls outside both regimes. For that manufacturer, the relevant question is rather what it must show when it presents simulation results to a vehicle authority, as Section 6.3 returns to.

> **Recommendation 1 (Provenance of action data).** Where a safety claim relies on what a world model predicts about actions, its developer should document how the actions in the training data were chosen, how they were recorded or inferred, and under what conditions they were observed, stating where any of these is unknown.
>
> **Recommendation 2 (Disclosure of shared dependencies).** Whoever relies on a world model to train or test a system should disclose any overlap, known or reasonably ascertainable, between the data or pretrained models used by the system and those used by the world model, where it bears on the safety claim.

[82] The proposal extends to action data an established practice of documenting how a dataset was composed and collected: Timnit Gebru and others, ‘Datasheets for Datasets’ (2021) 64(12) Communications of the ACM 86.

### 6.2. Accuracy in Control, Advisory, and Content World Models

A world model predicts the next state of an environment, and since it is typically run forward over several steps, each predicted state becomes the natural starting point for the next prediction. On this basis, control and advisory models compare possible actions, and a content model generates a scene that continues over time. In either case, because of this dependence between successive latent states, a mispredicted state can produce a compounding error.[83] Compounding errors are common in systems that generate step-by-step predictions, including LLMs, and that is relevant for rules like those in the AI Act (Arts 9(8), 15(1) and (3)) that ask providers of high-risk systems to test against "prior defined metrics and probabilistic thresholds" suited to the intended purpose, to ensure appropriate levels of accuracy and to declare them in the instructions for use. In the Chinese context, the standard on automated driving systems requires testing, including in simulation, while Article 4(5) of the Interim Measures only asks providers of public generative services to take effective measures to improve the accuracy of what they generate. The California regulation does not impose a general obligation to declare predicted accuracy, but prediction errors of world models might fall within its rules for assessing and mitigating catastrophic risk (§ 22757.12) only for large frontier developers, and only if they contribute to one of the catastrophic incidents the Act lists.

The main problem is that rules of this kind do not specify how far ahead the predictions must be checked against accuracy metrics, which is critical given that a model can be very accurate one step ahead and badly wrong five steps ahead. If only the first figure appears in the instructions for use, it can mislead the user about the reliability of the chosen action. Part of the reason this happens is, as already mentioned, that each prediction builds on the previous one, but the other part lies in what is generally called model exploitation. So, for example, a planner chooses some actions over others based on the model's predicted score for each action relative to the goal(s) the planner has been given (e.g., reaching a destination without collisions). The estimates always contain some error, and since the planner picks the highest score, it will tend to pick actions the model has overrated (and pass over the underrated ones). And the literature on world models describes this weakness from the model's perspective: since they are mostly trained on successful demonstrations, world models tend to predict success even for actions that would fail.[84] The two weaknesses add up as the planner selects the actions the model overrates, and the model overrates failing actions in particular. The very gap between the performance that a model predicts for a system optimized on it and the performance that the system then shows in reality, which is the direct measure of this problem, is almost never reported.[85] Such errors are normally exposed when a system is tested in the real world, but they become much harder to expose if the world model replaces real-world testing (more on this in Section 6.3). What is particularly interesting is that these errors not only make world models worse predictors but also affect the safeguards that rely on those predictions. In a recent experiment, a simulated robot arm planning with a world model was given a set of rules to follow (e.g., not to push an object into a forbidden area), and each move was checked against those rules based on the model's predictions before being executed. With perfect knowledge of the scene, the robot respected the rules 93% of the time, while relying on the world model it did so only 64.7% of the time, and more than half of the corrections imposed on it were for violations that had never happened.[86]

---

[83] Talvitie (n 48).

[84] Kamat and others (n 49) ss 6.5 and 6.8 and Table 23. For primary evidence, see T Yang and others, 'MiraBench: Evaluating Action-Conditioned Reliability in Robotic World Models' (arXiv preprint 2605.29360, 2026), where the failure is named 'optimism bias', and J Xu and others, 'World Models Dream of Success: Diagnosing and Repairing Failure Insensitivity in Robot World Models' (arXiv preprint 2610.09134, 2026), who report success-like predictions on verified failures in four released models.

[85] Y Yu and others, 'How Should World Models Be Evaluated for Embodied Decision-Making? A Decision-Making-Centric Position' (arXiv preprint 2606.15032, 2026) ss 3.6 and 5.

The way to fill this gap is, therefore, an adjustment in how rules of this kind are applied or interpreted, so that accuracy is measured and declared at relevant intervals over the same span of time on which the planner relies when choosing its actions (its *planning horizon*), tracking any deterioration or improvement as that span lengthens. And, for a related reason, accuracy should be checked for the actions the planner actually chooses, because average accuracy across all actions in a test set can give a misleading picture of error rates in practice. This proposal is grounded in recent proposals in the technical literature that point in the same direction: error should be reported as a function of how far ahead the model predicts. The model should be shown actions known to fail to check that it does not predict success for them (this technique is sometimes labeled optimism bias detection)[87], and predicted and real performance should be compared for the system actually optimized on the model (to measure what is sometimes called exploitability gap).[88]

A clarification is needed here: outcome-level measures take priority because predictive accuracy, even when reported over the whole horizon, is not enough on its own. It has been found not always to track the performance of the system that plans with the model, and a model with low prediction error can still rank two courses of action in the wrong order.[89] What a provider should therefore declare, first of all, is how well the model predicts the outcomes on which the safety claim rests, e.g., whether a maneuver ends in a collision, and how far the predicted performance for the system departs from the performance the system shows in reality. The latter is hard to measure at scale on real hardware, as its proponents acknowledge,[90] which is why it should be required in proportion to how much the safety claim rests on the model, as the EU vehicle rules already do for simulation (Section 5.1.2).

Since compounding errors and model exploitation are well-known in machine learning, this is a matter of detail in the standards for applying these rules, and the proposals just mentioned show what such standards, including those in preparation mentioned at the beginning of Section 5, could contain. In fact, the same literature suggests that whoever presents a world model should state in advance the tasks, the (planning) horizon, and the use for which it is offered, which, under the AI Act, would simply mean to specify the intended purpose that providers must already declare (Art 3(12)).

Finally, the same reasoning should extend from testing before approval to the system's use afterward (sometimes called post-market monitoring). A world model makes explicit predictions about what will happen next, which can later be compared with what actually happened in practice, so that the gap between the two offers a running measure of its accuracy in use, and a sign that the model is drifting outside the conditions in which it was tested.[91] The AI Act already requires high-risk systems to keep logs that support post-market monitoring (Art 12(1)–(2)), and providers to collect and analyze data on their performance throughout their lifetime, in a way that is “proportionate to the nature of the AI technologies” (Art 72(1)–(2)). For world models, this would mean recording the predicted outcome alongside the observed outcome and treating a growing gap between them as grounds for corrective action (Art 20(1)).[92]

[86] D Waldner and others, 'Legislating World-Model-Based Planning with Legal Reasoning' (arXiv preprint 2609.15113, 2026) s 6 (Questions 1, 4 and 5).

[87] Yang and others (n 86).

[88] Kamat and others (n 49); Yu and others (n 87).

[89] N Lambert and others, 'Objective Mismatch in Model-based Reinforcement Learning' in *Proceedings of the 2nd Conference on Learning for Dynamics and Control* (PMLR 120, 2020) 761; LM Bhamidipaty and others, 'Imperfect World Models are Exploitable' (arXiv preprint 2605.15960, 2026).

[90] Yu and others (n 87).

[91] A Sattar, 'Evaluating World Models: Why Traditional AI Benchmarks Fail' (Medium, 2026) <https://medium.com/@adnansattar09/evaluating-world-models-why-traditional-ai-benchmarks-fail-65777077dc07>.

> **Recommendation 3 (Accuracy for the intended use).** The developer of a world model should declare its error as a function of how far ahead it predicts, and the tasks and uses for which the figures hold. Whoever builds a system on it should declare, for the planning horizon and the actions the system actually chooses, how well the model predicts the outcomes on which the safety claim rests, and the gap between predicted and actual performance.
>
> **Recommendation 4 (Monitoring predictions in use).** Where outcomes can be observed, the system provider should compare a proportionate sample of forecasts with outcomes for the actions actually taken over the forecast period. It should specify in advance how discrepancies will be measured and when they trigger review.

### 6.3. Validation of Surrogate World Models

There is nothing inherently new about having AI systems trained or tested by other systems: e.g., LLMs can produce text that is then used to train further language models. Under some conditions, however, this reiterates and amplifies errors, as already observed.[93] As I pointed out in Section 2.4, a world model used to train or test another system (aka a surrogate model, Section3) produces synthetic data, something many of our legal systems were not prepared for because their data governance was designed with collected data in mind. Contrary to the latter, the reliability of synthetic data depends much more on the validity of the ground-truth assumptions used to produce them.[94] Along similar lines, the problem posed by world models concerns how a simulated environment for testing or training can be validated for that use, which requires checking its ground-truth assumptions, given that nobody wrote them.

Imagine that driving software is tested within a world model, so that the model generates a driving scenario and tests the software's behavior in that environment (with responses, always generated by the model, from the environment). Now imagine that the model failed to learn what the specific effect of a wet road is on the car's braking, e.g., in terms of how long it takes to stop. Although this generated scene abstractly contains elements that a person can inspect and verify, its visual plausibility is insufficient to establish that the car stops where a real car would, just as the plausibility of fluent text generated by an AI does not establish whether the text is correct.[95]

---

[92] Comparing forecasts with outcomes is established practice in United States banking supervision, where ‘outcomes analysis’, including back-testing at a frequency that matches the model’s forecast horizon, is part of model validation, and validation continues after a model goes into use: Board of Governors of the Federal Reserve System and Office of the Comptroller of the Currency, Supervisory Guidance on Model Risk Management (SR Letter 11-7, 4 April 2011) s V. The guidance was replaced in April 2026 by SR 26-2.

[93] I Shumailov and others, ‘AI Models Collapse When Trained on Recursively Generated Data’ (2024) 631 Nature 755.

[94] Gal and Lynskey (n 25); Cofone, Strandburg and Tilmes (n 25).

[95] This has also been tested: when several world models were put to work inside real tasks, their visual quality did not guarantee success, and what mattered more was whether they responded correctly to actions. See J Zhang and others, ‘World-in-World: World Models in a Closed-Loop World’ (International Conference on Learning Representations, 2026).

If for "conventional" AI and language models there is normally something outside the system against which outputs can be compared and verified, e.g., through real cases, an expert human reader, original data, or written equations, validating world models is more complex for a number of reasons. One of the primary reasons is that the output of the simulation, even when it takes the form of images or video, cannot be judged just by looking at it, as it describes a complex physical scenario, sometimes merely hypothetical, e.g., what would physically happen after the tested system acts and in correspondence with a further potential action. This, in the first place, and as already noted,[96] calls for new benchmarks capable of certifying whether a world model's learned internal environment closely enough matches reality to guide a real-world decision. On this aspect, the technical literature has started to say what such benchmarks should measure: not how real a scene looks, but, e.g., whether the prediction changes as reality would, especially when, from the same starting scene, the action changes ("action following" or "controllability"); whether quantities that have a physical truth, like speeds and distances, come out right.[97]

These tests require that reality can be consulted, although the extent of real-world evidence needed depends on the claim and the use of a world model.[98] So, for instance, a manufacturer need not recreate a near-collision to check the model's prediction of stopping distance because braking measurements taken on a clear track, under comparable conditions, may suffice.[99] Some relationships can also be checked without knowing the exact outcome of either run: for example, if a vehicle continues at a known constant speed for one additional second before applying the same brakes on a uniform surface, the additional distance traveled during that second can be calculated independently. But the difficulty is greatest, and so we get to the second set of reasons that make validation tricky, where world models are more useful: to create situations that are rare in real data or too dangerous or costly to run in practice.[100] On top of that, a world model does not give only one answer but a whole chain of future moments, some of them action-conditioned, which makes the number of situations it can end up in too large to test.

These difficulties kick in particularly when we read them in combination with the rules like those in the AI Act, which require training, validation and testing data to be representative of the intended use, and data governance practices that concern the assumptions behind the data, an assessment of their availability and suitability, and the identification of relevant gaps and how they can be addressed (Art 10(2)(d), (e) and (h), (3)–(4)). The Act also defines testing data as data providing "an independent evaluation" of the system (Art 3(32)). In machine-learning practice, an evaluation is independent when it is run on data kept apart from those used for training.[101] Both requirements, though, were written for data that are collected, and they become hard to apply when the data are generated. Representativeness is ordinarily assessed by comparing the data with real cases, but in the rare situations in which world models are used to produce them, such cases are scarce or absent. Independence is ordinarily ensured instead by keeping two sets of data apart, but two sets generated by the same world model, or by models that learned from the same records, remain dependent, however carefully they are separated, since they share the same assumptions and the same gaps. In other words, when data are generated, what makes them representative and independent depends on the environment that generates them, and this is where the gap lies. The problem is that the Act does not cover that environment and mentions synthetic data only as a substitute for real data in detecting bias and in sandboxes (Arts 4a(1)(a) and 59), but never states when data of this type can count as evidence. In the Act and elsewhere, the rules that mention synthetic data focus on its origin but not on its validity: e.g., the Commission's template for general-purpose models asks if synthetic data were used and which models generated them (see Section 5.1.1), California mandates developers to disclose the use of synthetic data (see Section 5.2), and both the Act and the Chinese Labeling Measures require generated content to be marked (see Sections 5.1.1 and 5.3).

---

[96]Stanford HAI (n 3).

[97]Zhang and others (n 97); Kamat and others (n 49); Yu and others (n 87).

[98]Scaling the evidence required of a simulation to the reliance placed on it is the approach taken by the United States Food and Drug Administration to computational models in medical-device submissions, where the credibility required depends on 'model risk', i.e., the model's influence on the decision combined with the consequence of a wrong decision: Food and Drug Administration, *Assessing the Credibility of Computational Modeling and Simulation in Medical Device Submissions: Guidance for Industry and Food and Drug Administration Staff* (November 2023), which builds on ASME V&V 40-2018 and applies to physics-based or mechanistic models.

[99]Nuclear safety regulation in the United States proceeds in a similar way for accidents that cannot be staged: the phenomena on which the result depends are identified and ranked, and the simulation model must represent each important one: US Nuclear Regulatory Commission, *Regulatory Guide 1.203: Transient and Accident Analysis Methods* (December 2005).

[100]Comparing the output of a model with reality, for the use it is meant for, is what the literature on simulation calls operational validation, and that literature distinguishes the cases in which data on the real system can be collected from those in which they cannot, where only weaker forms of validation remain: RG Sargent, 'Verification and Validation of Simulation Models' (2013) 7 Journal of Simulation 12. The rare situations just mentioned belong to the second group.

How, then, should this legislative silence be read? The analysis above suggests that it may be the source of a restrictive interpretation, which may also be a way to fill the gap: i.e., generated evidence should count in proportion to how far it can be checked against real data. For the same reason, a test should not be deemed independent if it is run in the same environment in which the system was trained, and simulated scenes of situations for which no real data exist should not, on their own, prove anything.[102] Where the system and the environment share only data or pretrained components, the matter is primarily one of degree, as most systems draw on some of the same public datasets and pretrained models. In these cases, independence should probably be presumed lacking only where what is shared bears on the behavior being tested, and the presumption should yield to a showing that the test still catches the relevant errors, for which Recommendation 7 offers one method. The concern is not abstract and has empirical support, although in a neighboring field, i.e., LLMs used as evaluators have been found to favor similar models.[103]

The biggest limitation here is that the Act's rules apply directly only to the products in the first list of Annex I and to the uses in Annex III, where the use of world models as test environments is not documented. For vehicles and machinery, which are the most common fields for world model testing, the rules apply only under sectoral law. The question, therefore, is whether the same reading finds support in the vehicle rules.

In this sense, the EU vehicle legislation includes some safeguards, although it is unclear how they would apply to learned simulators like world models. According to the current rules, the manufacturer must provide information on the data and scenarios used to validate the simulation, and these data must sufficiently reflect the vehicle's operating domain. Therefore, the outcome of the credibility assessment defines the envelope within which the simulation can be used to support the approval (Annex III, Part 4). Outside that envelope, simulation results cannot be validated and support the approval.

---

[101] T Hastie, R Tibshirani and J Friedman, *The Elements of Statistical Learning* (2nd edn, Springer 2009) ch 7.

[102] The independence of those who check a system from those who built it is defined along technical, managerial and financial lines in IEEE Std 1012-2016, Standard for System, Software, and Hardware Verification and Validation; United States banking supervisors likewise expect validation to be carried out by staff who are not responsible for developing or using the model: SR Letter 11-7 (n 94) s V.

[103] S Goel and others, 'Great Models Think Alike and this Undermines AI Oversight' in *Proceedings of the 42nd International Conference on Machine Learning* (PMLR 267, 2025) 19621.

Here as well, though, there is some room for legal vagueness and potentially for a gap. First, the envelope rests on the data used to validate the simulation against reality, while the Regulation says nothing about the data the simulator was trained on, which, for a world model, determines where it can be expected to work at all. Second, the Regulation does not say how the envelope is to be drawn, which is equally important: a world model may have been checked on dry roads at any speed and on wet roads at low speed, yet never at high speed on a wet road. Listing the conditions separately would hide that gap, since "high speed" and "wet road" would both appear as checked, and a model that learns from examples may badly mispredict a combination it has never seen. The envelope should therefore list only the combinations of conditions that were actually checked against real driving and that the model encountered while learning.

Moreover, there is a tension within the Regulation itself, which allows simulation to be used "in particular for scenarios that are difficult on a test track or in real driving conditions" (Annex III, Part 2). Now, if for a conventional simulator, the equations validated in ordinary conditions offer some basis, albeit itself debated, for applying it to less ordinary ones[104], for a learned simulator, such basis is much more controversial, since its behavior beyond what it was trained and validated on is exactly what the studies above call into question. In other words, the Regulation encourages simulation where a learned simulator is least reliable, without saying how its envelope is to be established there. This calls for an adjustment, inspired by the same principle I mentioned with reference to the AI Act: the envelope should be drawn as described above, and the encouragement to simulate hard scenarios should not extend beyond it.

Both readings, however, still leave open how to judge a test precisely where real data are missing, i.e., where the envelope ends. A technique that could help there, and that is not often (if not never, to my knowledge) considered in legal documents, is *mutation testing*. In software engineering, the quality of a set of tests is assessed by intentionally introducing small defects in the program and verifying whether the tests catch them, a practice that has also been adapted to machine-learning systems.[105] The proposal, mentioned in Section 6.2, of showing a world model actions that are known to fail follows the same logic, and it can be extended from the model to the whole test: versions of the tested system with a known defect, e.g., driving software that brakes too late on a wet road, are run through the same test inside the world model, and if they pass, the test has been shown unable to catch that kind of defect. In the example above, the omission of the wet road would come to light without any real recording. We should be aware, though, that the technique has some limits: it only reveals blindness to the defects someone thought to plant, leaving outside the unexpected ones, and a defective version that fails proves little, as its failure is itself generated by the model. Its legal footing, however, is already partly there. Under the AI Act, testing must "ensure" that a high-risk system performs consistently for its intended purpose, against "prior defined metrics and probabilistic thresholds" (Art 9(6) and (8)), and a test unable to catch the defects that matter can hardly ensure it; the share of planted defects that a test catches could serve as one of those metrics, and be reported among the "validation and testing procedures" in the technical documentation (Annex IV, point 2(g)). The Act's only reference to adversarial testing, i.e., red teaming, shares the same attitude of looking for failures rather than confirming success, but it concerns general-purpose models with systemic risk and the model itself, not the adequacy of a test (Art 55(1)(a) and Annex XI, Section 2). Like the tests discussed above, mutation testing could therefore enter the existing frameworks through the standards and guidance that apply the AI Act, as a way of carrying out the credibility assessment that the EU vehicle rules already require, and as one of the validation methods that a manufacturer must describe in its safety case in California, while a new rule would be needed only to impose it on the supplier of the world model, on whom none of the frameworks examined places a duty of this kind.[106]

---

[104] Oberkampf and Roy (n 15). On the limits of conventional simulation for this purpose, see A Stocco, B Pulfer and P Tonella, 'Mind the Gap! A Study on the Transferability of Virtual Versus Physical-World Testing of Autonomous Driving Systems' (2022) IEEE Transactions on Software Engineering <https://doi.org/10.1109/TSE.2022.3202311>, who compare a small physical vehicle with its digital twin and find that many test results transfer from one to the other, but not all.

[105] Y Jia and M Harman, 'An Analysis and Survey of the Development of Mutation Testing' (2011) 37 IEEE Transactions on Software Engineering 649; L Ma and others, 'DeepMutation: Mutation Testing of Deep Learning Systems' (IEEE International Symposium on Software Reliability Engineering, 2018).

The same questions arise for the Chinese standards on automated driving systems, while the California regulation is much more underdetermined: i.e., the developer's own assessments determine whether a frontier model could pose a catastrophic risk (§ 22757.12(a)(2)–(5), (c)(2)), and leaves their method to the developer.

> **Recommendation 5 (Evidentiary weight of simulated results).** Evidence generated by a world model should support a safety claim in proportion to how far the conditions involved, or their components, have been checked against real data and are covered by the model's training data. Where they have not, generated scenes may be used to discover situations in which the system could fail, to be investigated and, where confirmed, corrected, but cannot on their own show that the system will not fail.
>
> **Recommendation 6 (Independence of validation evidence).** A test should not be regarded as independent when it is run in the environment in which the tested system was trained. Where the two share only data or pretrained components relevant to the tested behavior, independence should be presumed absent unless the test is shown to detect the relevant errors.
>
> **Recommendation 7 (Detection of safety-relevant failures).** Those who rely on tests run inside a world model should show that the test is able to fail a defective system, e.g., by running versions of the system with known defects, drawn from its hazard analysis or a standard, through the same test. They should report any missed defect and its implications for the claim. Detecting the defects establishes that the test can catch them under those conditions, but not that the test is enough to show the system is safe (except in the conditions to which Recommendation 5 gives weight).

## 7. Conclusion

My main question in this article was about whether AI regulations are ready for world models. For this purpose, I used three of the most developed legal contexts on the topic - EU, California, and China - and showed that they can all virtually reach them, but with important gaps that should be filled. The main shortcoming I found concerns world models used to train or test other systems, where the evidence of safety is itself produced by a model, and where the current rules on data, designed for data that are "conventionally" collected, can be extended to cover where generated data come from but not when they can be trusted. More generally, the analysis suggests that governing world models is, at bottom, a matter of governing a chain of evidence.

I would like to emphasize one last consideration that goes beyond the rules and concerns the likelihood that world models will extend the competence required for governance. If the governance schemes and bodies that AI regulations have set up were built around a kind of expertise - e.g., the AI Act asks national authorities for competence in AI technologies, data, cybersecurity, fundamental rights, and health and safety risks (Art 70(3)) - this may not be enough anymore. Assessing and governing world models means judging whether a simulation is credible, whether a predicted motion is physically correct, and how a machine behaves when it acts on its predictions, which are matters of modeling, physics, and robotics more than of data and software alone.

[106] A precedent for attaching requirements to a tool, and not only to the product, exists in automotive functional safety, where the confidence needed in a software tool depends on whether its malfunction could introduce errors or fail to detect them, and on how likely such errors are to be caught otherwise: ISO 26262-8:2018, Road vehicles — Functional safety — Part 8 cl 11.

Let me conclude by flagging that these conclusions I reached in this article remain provisional, as the technology is immature and most of what is known about its use comes from the companies that build it.